\documentclass[pdflatex,sn-aps]{sn-jnl}

\usepackage{multirow}%
\usepackage{mathrsfs}%
\usepackage[title]{appendix}%
\usepackage{textcomp}%
\usepackage{manyfoot}%
\usepackage{booktabs}%
\usepackage{listings}%

\usepackage{graphicx}
\usepackage{dcolumn}
\usepackage{amsmath}
\usepackage{amssymb}
\usepackage{amsfonts}%
\usepackage{float}
\usepackage{subfigure}
\usepackage[dvipsnames]{xcolor}
\usepackage{algorithm}
\usepackage{algorithmicx}%
\usepackage{algpseudocode}
\usepackage{pgfplots}
\pgfplotsset{compat=1.18}
\usepackage{bm}
\usepackage{epstopdf}
\usepackage{amsthm}
\usepackage{soul}
\graphicspath{{./figures/}}
 \def\V{{\mathrm{V}}}
\def\U{{\mathrm{U}}}
\def\Q{{\mathrm{Q}}}
\def\P{{\mathrm{P}}}
\def\R{{\mathrm{R}}}
\def\A{{\mathrm{A}}}
\def\C{{\mathrm{C}}}
\def\W{{\mathrm{W}}}
\def\S{{\mathrm{S}}}
\newcommand{\N}{ \mathbf{N} }

\newcommand{\F}{\mathbf{F}}

\newcommand{\x}{ \mathbf{x} }
\newcommand{\y}{\mathbf{y}}
\newcommand{\X}{ \mathbf{X} }

\newcommand{\tr}{^\intercal}

\newcommand{\coupling}{ \bm{\varsigma} }
\newcommand{\stabilizer}{ \bm{\chi} }
\newcommand{\jacobian}{\nabla_\x\tr}

\def\prl#1#2#3{{ Phys. Rev. Lett.} {\bf #1}, #2 (#3)}

\def\pre#1#2#3{{ Phys. Rev. E} {\bf #1}, #2 (#3)}

\def\beq{\begin{equation}}
\def\eqn{\end{equation}}

\def\prl#1#2#3{{ Phys. Rev. Lett.} {\bf #1}, #2 (#3)}

\def\pre#1#2#3{{ Phys. Rev. E} {\bf #1}, #2 (#3)}

\theoremstyle{thmstyleone}%
\theoremstyle{thmstyletwo}%

\theoremstyle{thmstylethree}%

\begin{document}

\title{Control Strategy for Generalized Synchrony in Coupled Dynamical Systems}

\author[1]{\fnm{Vishal} \sur{Juneja}} 
\author[2]{\fnm{Suresh} \sur{Kumarasamy}} 
\author[3]{\fnm{Aryan} \sur{Patel}} 
\author[4]{\fnm{Amrita} \sur{Punnavajhala}} 
\author[5]{\fnm{Ram} \sur{Ramaswamy}}\email{rramaswamy@gmail.com}
\affil[1]{\orgdiv{Department of Geophysics}, \orgname{Institute of Science, Banaras Hindu University}, 
\city{Varanasi}, \postcode{221005}, \state{Uttar Pradesh}, \country{India}}
\affil[2]{\orgdiv{Centre for Computational Modeling}, \orgname{Chennai Institute of Technology}, \city{Chennai}, 
\postcode{600069}, \state{Tamil Nadu}, \country{India}}
\affil[3]{\orgname{Raphe mPhibr Pvt. Ltd.}, \city{Noida}, 
\postcode{201305}, \state{Uttar Pradesh}, \country{India}}
\affil[4]{\orgdiv{Department of Applied Mathematics}, \orgname{University of Waterloo}, \city{Waterloo}, 
\postcode{N2L 3G1}, \state{Ontario}, \country{Canada}}
\affil[5]{\orgdiv{Department of Physical Sciences}, \orgname{Indian Institute of Science Education and Research}, 
\city{Berhampur}, \postcode{760003}, \state{Odisha}, \country{India}}

\abstract{Dynamical systems can be coupled in a manner that is {\em designed} to drive 
the resulting dynamics onto a specified lower dimensional submanifold in the 
phase space of the combined system. On the submanifold, the variables of the 
two systems have a well-specified functional relationship. This 
process can be viewed as a control technique that ensures {\em generalized 
synchronization}. Depending on the nature of the dynamical systems and the 
specified submanifold, different coupling functions can be derived in order 
to achieve a desired control objective. We discuss a circuit implementation 
of this strategy for coupled chaotic Lorenz oscillators, as well as a demonstration
of the methodology for designing coordinated motion (swarming) in a set of 
autonomous drones.}




\begingroup
\renewcommand\thefootnote{}
\footnotetext{{\large Accepted for publication in EPJ Special Topics (EPJ ST).}}
\addtocounter{footnote}{-1}
\endgroup

\maketitle

\section{Introduction}

The application of control theory to nonlinear dynamical systems \cite{control} and the 
study of synchronization phenomena in chaotic systems \cite{pc1990,yf1983,av1986,PRK, 
SHS, CD,pc2015} are research areas practical importance for over three decades. These have been 
developed more or less in parallel, with many synchronisation methods being cast as control techniques. 
The reverse is less common, since control objectives need not always correspond to specific 
dynamical outcomes. \\

In the present paper we discuss a situation where the correspondence works in both directions. 
We couple two dynamical systems in such a manner that the collective 
dynamics is confined to a {\em specific} submanifold in the phase-space 
of the coupled system. This is the required control objective, and it
is equivalent to the generalized synchronization of the coupled dynamical 
systems. Recall that non-identical systems are said to be in generalized
synchrony when the  variables of the individual systems become functionally 
related \cite{RSTA, ARS, HOY, pp2008, CR,kjr,unr}. This functional 
relationship specifies the submanifold in the phase space of the combined 
system \cite{KJ}. Thus geometric control objectives can clearly 
be seen as a means of {\em designing} generalized synchronization (GS)\cite{CR}.\\

The control objective is equivalent to constraining the dynamics through the design of 
suitable coupling functions. Our approach \cite{CR} involves the solution of a set of 
under-determined equations, so there is 
considerable choice in the forms of the control or coupling terms that will give the desired result. This flexibility
makes the process both adaptive and robust; even for the case of perfect synchrony in identical systems when all the 
variables coincide and the synchronous motion occurs on the so--called synchronization manifold, there are a variety 
of different couplings that can be utilized. The possibility of synchronizing two or more chaotic
systems in this manner has inspired a large body of work in areas ranging from secure communication and chaos control 
\cite{co1993, KEK} to synthetic biology \cite{sb} and the study of electrical power grids \cite{grids,grid2}, 
making the study of GS in complex systems an area of considerable experimental and 
theoretical importance.\\

We demonstrate a practical implementation of the control method for the 
GS of two electronic circuits that model the Lorenz oscillator \cite{l1963}.
Practical implementation of different chaotic synchronization techniques has, 
from the start, been explored in electronic  circuits \cite{cp1991,co1993}. 
In addition to providing a physical 
realization of many abstract dynamical systems, circuit experiments help probe 
the validity and robustness of control techniques. In addition, novel chaotic systems 
have also been devised first as circuits, with the equations of motion 
being studied in depth only subsequently \cite{Chua}.\\

Reverse-engineering approaches to synchronization have been devised in the past in various different 
contexts \cite{pp2008,gp2008,gb2009,pd2010}. Some of these applications, such as those using the OPCL 
(or open plus closed loop) coupling are highly stable, but are limited in the kinds of states that 
can be targeted \cite{JG}. Projective synchronization  \cite{MR} and its generalizations \cite{GH} 
have also been a topic of considerable interest, and there is some overlap in the procedures employed 
in generalized projective synchrony and the present approach. However, there are important differences, 
primarily to do with the flexibility in the design principles that are inherent in the present control technique. \\

Below we review the basic principles of our coupling strategy. Details of the circuit implementation 
for specific cases are discussed in Section 3 where we present experimental results on 
coupled circuits.  Implementaion of the algorithm in a set of drones is described and analysed in 
Section 4. We conclude with a discussion and summary in Section 5. 

\section{Control onto a desired submanifold}

The general methodology that was proposed earlier \cite{CR} can be viewed as a geometric control technique.
Since the objective is to constrain the dynamics of the coupled  system to a specific submanifold in the phase space, 
the defining equations of this hypersurface are expressed as algebraic relations between the variables
of the two systems, namely as a set of constraints. This gives, via a straightforward procedure, to a set 
of requirements for the coupling between the two systems. There is flexibility in the choice of coupling function; 
as is well known, the same form of synchronization can be achieved with a number of different couplings. We summarise 
the main equations below.\\

Consider two independent systems, with variables $ \x \in \mathbb{R}^{m}$ and $\y\in\mathbb{R}^{n}$ with flows 
specified by the functions $\F_1(\x)$ and $\F_2(\y)$ respectively. 
The aim is to couple them suitably so that the resulting dynamics satisfies the conditions
\begin{equation}
    \begin{aligned}
    \label{eq1}
\y=\Phi[\x] 
\end{aligned}
\end{equation}
which is a functional relationship between the variables of the two systems. (A more general functional relationship between
the systems could be non-separable, given for example by the condition $\Phi[\x,\y] = 0$.) When coupled, the equations of 
motion become
\begin{equation}
    \begin{aligned}
    \label{eq2}
    \dot{\x} &= \F_1(\x) +\epsilon \coupling_1(\x,\y) \\
    \dot{\y} &= \F_2(\y)+\epsilon \coupling_2(\x,\y).
\end{aligned}
\end{equation}
where $\coupling_i$'s are coupling terms that need to be determined such that the dynamics obeys the condition Eq.~(\ref{eq1}) 
and $\epsilon$ is the strength of the coupling. We had not explicitly included the coupling constant in our earlier work \cite{CR}
since the algebraic form of the coupling function does not depend on it.  For simplicity we have taken both coupling
terms to have the same strength of coupling; clearly this can be generalised. 
We can rewrite Eq.~(\ref{eq2}) compactly by introducing the notation 
$\X \equiv  [\x~\y]^{\intercal} \in \mathbb{R}^{m+n}$, 
$\F(\X) \equiv [\F_1(\x)~\F_2(\y)]^{\intercal}$ and 
$\coupling(\X) \equiv [\coupling_1(\x,\y)~\coupling_2(\x,\y)]^{\intercal}$. 
This gives
\begin{equation}
    \dot{\X}= \F(\X) + \epsilon\coupling(\X),
    \label{eq3}
\end{equation}
namely as a dynamical system in a phase space of dimension $m+n$. 
The motion in the combined system is to be confined to a lower-dimensional subspace $ \mathcal{M}$ that is specified by 
a set of  $N < n+m$ functional relations between the variables of the two systems, namely the condition
\begin{equation}
\label{constraint}
\Phi(\X) = [\phi_1 (\x,\y) \ldots \phi_N (\x,\y)]^{\intercal} = 0, 
\end{equation}
which are the required set of constraints. In order to bring
the dynamics onto the submanifold, our basic strategy is to ensure that the flow of the combined system is 
orthogonal to the normals to the submanifold. In each of the directions in phase space, these are given by
\begin{eqnarray}
    \N_i(\X) &=& \nabla_{\X} \phi_i (\x,\y), ~~~i=1,\ldots, N,
 \end{eqnarray} 
and collectively they give the matrix of normals
\begin{eqnarray}
    \mathfrak{N} &\equiv& \jacobian \Phi(\X)  = \begin{bmatrix} \N_1 & \N_2 & \cdots \N_N \end{bmatrix} \tr.
\end{eqnarray}
In the coupled system, the flow is orthogonal to the normals, and this gives the condition
\begin{equation} \label{flowBalancing}
\epsilon\mathfrak{N} \coupling = - \mathfrak{N} \F,
\end{equation}
from which the coupling functions $\coupling_i$ can be determined. \\

{\color{red}{As we have noted earlier \cite{CR}, additional terms 
can be added into the coupling to {\em stabilize} the control, the only requirement
being that they should vanish on the specified  manifold. Clearly, setting $ \coupling(\X) \to 
\coupling(\X)+\stabilizer(\X)$ in Eq.~(\ref{eq3}), where the term $\stabilizer(\X)$  
takes the value zero on the target manifold will still achieve control. A suitably chosen
$\stabilizer(\X)$, for instance taking it to be a Lyapunov function  \cite{sontag}, provides an efficient 
strategy for ensuring stability.}} \\

In the so--called ``master-slave" scenario, where $\coupling_1(\x,\y) = 0$, a simpler 
strategy is available. As one can see easily, the choice of 
\begin{equation}
    \coupling_2(\x,\y) = -  \F_2(\y)+\epsilon (\Phi[\x]-\y)+  \nabla_{\x} \Phi[\x] \F_1(\x)
\end{equation}
leads to the following condition,
\begin{eqnarray}
   \frac{d}{dt}(\Phi[\x]-\y) &=& \nabla_{\x} \Phi[\x]\cdot \F_1(\x) - \F_2(\y) - \coupling_2(\x,\y)\\
   &=&-\epsilon (\Phi[\x]-\y) .
   \label{errorDynamics}
\end{eqnarray}
Thus $ (\Phi[\x]-\y)$ will decay exponentially, leading directly to the required condition, 
namely Eq.~(\ref{eq1}). Note however, that in this scenario,
the dynamics of the master system in $\x$ is unchanged while that of the slave system $\y$ becomes 
\begin{eqnarray}
\dot{\y} &=&  \epsilon (\Phi[\x]-\y) +   \nabla_{\x} \Phi[\x]\cdot \F_1(\x) 
\end{eqnarray}
so that the intrinsic dynamics of the slave system is completely suppressed. Such an approach, while not within the 
usual framework of synchronization, is useful when considering a constraint that is translational, namely when 
\begin{equation}
\Phi[\x] = \x-\mathbf{c},
\end{equation}
with $\mathbf{c}$ being a vector of constants so that the desired manifold is specified by conditions of the form 
\begin{equation}
y_i=x_i-c_i. 
\label{lyapfun}
\end{equation}
In that case, the equation of motion for the slave variables becomes 
\begin{eqnarray}
\label{translationalContraintMainEqn}
\dot{\y} &=&   \F_1(\x) + \epsilon (\x-\y-\mathbf{c}).  
\end{eqnarray}
As we show below, this equation is useful in dealing with spatially separated systems, 
{\color{red}{and as can be seen, quadratic Lyapunov functions \cite{Massera} that ensure that Eq.~(\ref{lyapfun})
is maintained can be added to the control.}} 
In Section \ref{swarmDronesSectionTitle}, we implement a swarm algorithm based on
Eq.~(\ref{translationalContraintMainEqn}).
\newline
\section{Applications}
We consider the coupling of two Lorenz oscillators  since the corresponding electronic circuits 
can be constructed in a fairly standard manner \cite{lorenz_circuit}. 
The flow equations are \cite{l1963}
\begin{eqnarray} 
    \dot{x}_1&=&\sigma_x (x_2  - x_1) \nonumber\\
    \dot{x}_2&=&(\rho_x  - x_3)x_1   -x_2   \nonumber\\
    \dot{x}_3&=&x_1 x_2  - \beta_x x_3 \label{lorenz_eqn}
\end{eqnarray}
for the $\x$ subsystem, and similarly for $\y\equiv (y_1, y_2, y_3)$ subsystem with parameters 
$\sigma_y, \rho_y, \beta_y$. The phase space of the combined system is thus six-dimensional. \\

\begin{figure}[h]
\centering
    \includegraphics[scale=0.8]{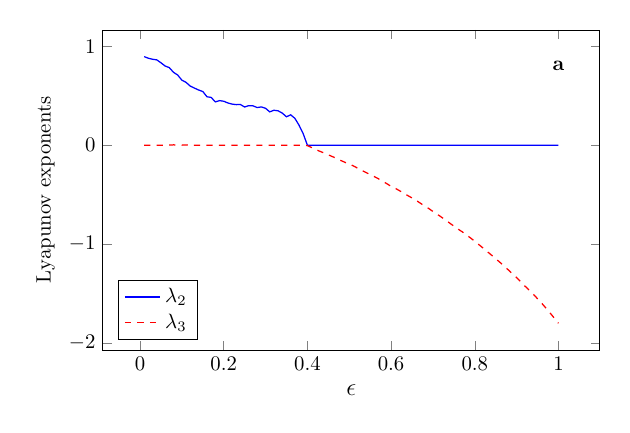}
    \includegraphics[scale=0.8]{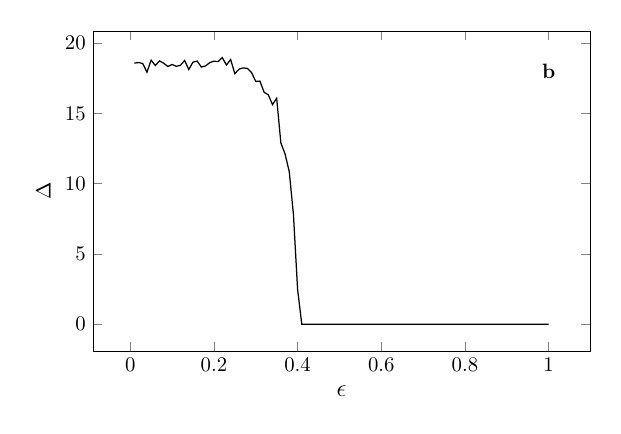}
\caption{(Colour online) Transition to complete synchronization as a function of coupling strength $\epsilon$ 
in the coupled Lorenz system; see Eq.~(\ref{pecora_carroll}). 
(a) The two largest transverse Lyapunov exponents, and (b) the order parameter $\Delta$ that measures deviations 
from the synchronization submanifold. The case discussed in \cite{pc1990} was for $\epsilon$=1.} \label{fig:pecora_carroll_LE}
\end{figure}

As was shown by Pecora and Caroll \cite{pc1990}, for the case when the parameters of both subsystems are 
identical, making one (say $\x$) the master and $\y$ the slave leads to complete synchronization on 
a three-dimensional subspace of the phase space. This is the synchronization manifold defined by 
three independent conditions (or constraints) $x_i-y_i=0, i=1,2,3$. In the present notation, the 
relevant coupling functions are
\begin{align} \label{pecora_carroll}
\coupling_1 = \begin{bmatrix}
0\\
0\\
0
\end{bmatrix}
\quad
\coupling_2 = \epsilon\begin{bmatrix}
0\\
(\rho - y_3) (x_1 - y_1) \\
(x_1 - y_1) y_2
\end{bmatrix},
\end{align}
with $\epsilon$ set to unity. Since $\coupling_1$ is a null-vector, the coupled equations have
a skew-product form with the dynamics of $\x$ (the master) unaffected by $\y$ (the slave)
 subsystem. The dynamics can be studied as a function of $\epsilon$ and the above coupling and we 
find that complete synchronization between the two systems is actually achieved for $\epsilon$ above 0.41.
In Fig.~\ref{fig:pecora_carroll_LE} the largest two transverse Lyapunov exponents of the coupled system
are shown as a function of $\epsilon$. The time-averaged distance of the coupled dynamics 
from the synchronization submanifold, namely
\begin{equation} \Delta = \langle\left\lVert{\y-\Phi[\x] }\right\rVert\rangle\label{op}\end{equation}
where $\langle\cdot\rangle$ denotes the time average is an alternate 
indicator of the synchronization. 
This quantity captures the somewhat abrupt nature of the transition, as can be seen in Fig.~\ref{fig:pecora_carroll_LE} (b).
\newline
\noindent {\bf Projective Synchrony: }
 \label{PS}
Any linear transformation of the synchronization manifold leads to projective synchronization \cite{MR}, namely when
 \begin{align}
   \begin{bmatrix}
 y_1\\  y_2 \\ y_3
\end{bmatrix}=  \cal{A}\cdot\begin{bmatrix}  x_1\\  x_2 \\ x_3
\end{bmatrix}
\end{align}
and $\cal{A}$ here is a 3$\times$3 matrix \cite{CR}. When the elements of $\cal{A}$, denoted $a_{ij}$, are such that
$a_{ij}=\alpha_i\delta_{ij}$, namely $\cal{A}$ is diagonal, one has the simplest case that corresponds
to a scaling of the variables. Both a master--slave type coupling
\begin{align} \label{linear_coupling}
\coupling_1 &= \epsilon \begin{bmatrix}
0\\
0\\
0
\end{bmatrix}, \nonumber \\ 
\coupling_2 &= \epsilon \begin{bmatrix}
\sigma_x \alpha_1 (x_2 - x_1) - \sigma_y (y_2 - y_1) + (\alpha_1 x_1 - y_1)\\
\alpha_2 x_1 (\rho_x - x_3) - (\rho_y - y_3) y_1\\
\alpha_3 (x_1 x_2 - \beta_x x_3) - (y_1 y_2 - \beta_y y_3) + (\alpha_3 x_3 - y_3)
\end{bmatrix}
\end{align}
\noindent
or bidirectional coupling:
\begin{align} \label{bi-directional}
\coupling_1 = \epsilon\begin{bmatrix}
\sigma y_2 / \alpha_1  \\
( \rho_y y_1 - y_1 y_3 )/\alpha_2 \\
y_1 y_2 / \alpha_3
\end{bmatrix}
 \quad
\coupling_2 = \epsilon\begin{bmatrix}
\sigma \alpha_1 x_2 \\
\rho_x \alpha_2 x_1 - \alpha_2 x_1 x_3 \\
\alpha_3 x_1 x_2
\end{bmatrix}
\end{align}
can be derived quite simply, and both of are effective in ensuring that the dynamics is on the desired submanifold. Note that 
the parameters $\rho_x$ and $\rho_y$ of the two subsystems need not be identical. For arbitrary values of the 
$\alpha_i$'s, this coupling ensures that the dynamics is on the desired projective synchronization manifold. Of course
when all $\alpha_i$ = 1 the systems are {\em completely synchronized} even though the system parameters can be different. 
The unidirectional coupling is essentially equivalent to the case studied by Pecora and Caroll
\cite{pc1990}, namely Eq.~(16),
but the bidirectional coupling leading to complete synchronization is a new scenario.\\

\begin{figure}[H]
    \centering
    \includegraphics[scale=0.04]{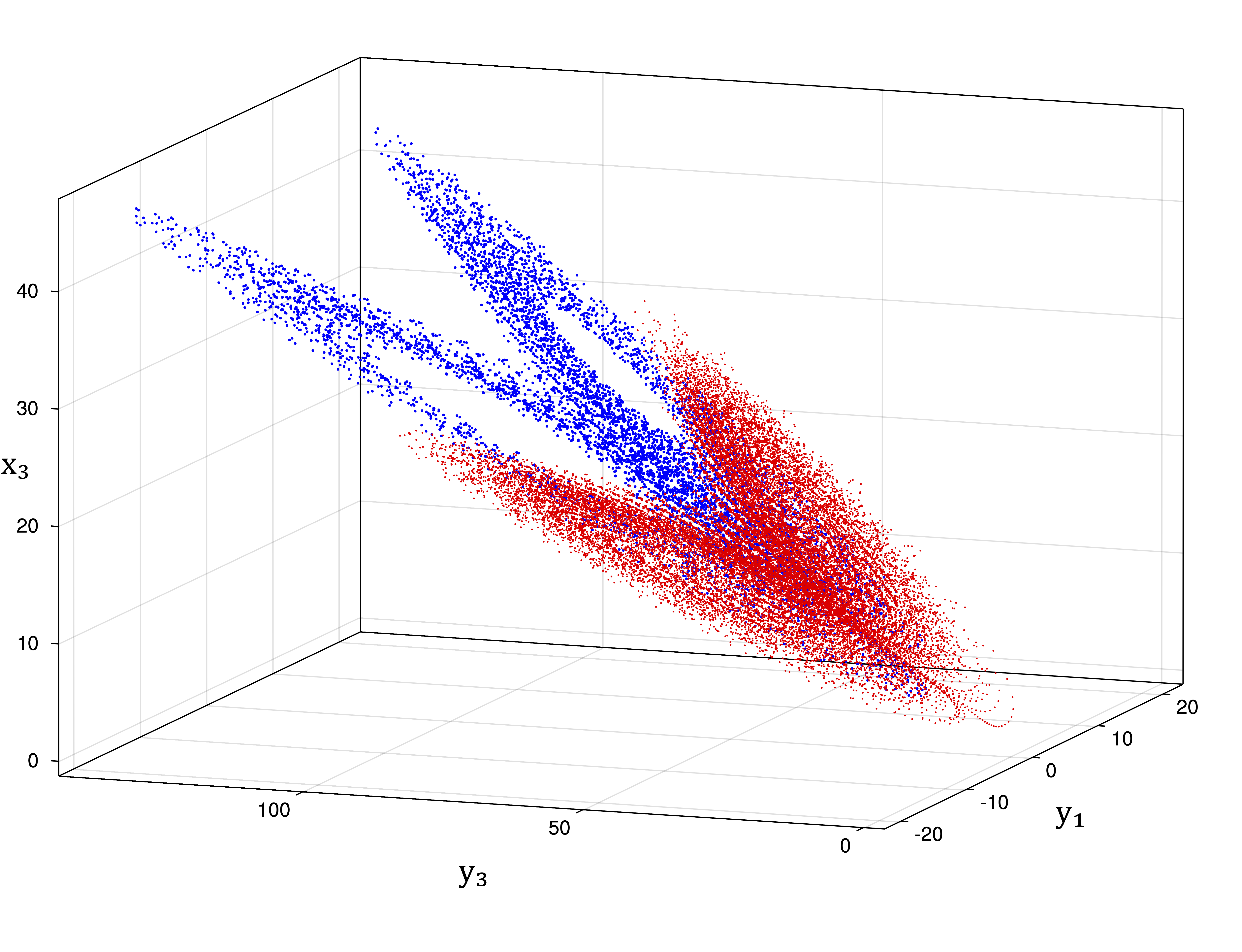}
        \caption{(Colour online) Projective synchronisation with $\alpha_1 = 1$, $\alpha_2 = 2$ and $\alpha_3 = 3$. 
        The blue dots are for unidirectional (master-slave) coupling while red dots show the dynamics with 
        bi-directional coupling. While the dynamics in either case is confined to the same plane,  the trajectories
        occupy different parts of the specified submanifold.  Here $\epsilon = 1$, but 
        the dynamics reaches the submanifold for smaller $\epsilon$ in both coupling cases. }
    \label{fig:bidirection}
\end{figure}

Choosing $\alpha_k = k$  for $k$=1-3 gives the results shown in Fig.~\ref{fig:bidirection} 
in which the coupled dynamics is projected on the plane specified by $kx_k=y_k$, with
master-slave coupling (blue) and bi-directional coupling (red). 
\noindent{\bf Nonlinear Projection:}
Our method applies quite easily to situations where the desired functional dependence is polynomial.
Since on the Lorenz attractor the variables $x_3$ or $y_3$ are always positive, as an illustration 
of our method we choose the constraint
\begin{align}
 \begin{bmatrix}
 y_1\\  y_2 \\ y_3
\end{bmatrix}=  \begin{bmatrix}  x_1\\  x_2 \\ x_3^2
\end{bmatrix}.
\end{align}
that retains the qualitative features of the dynamics, while targeting the dynamics
onto a submanifold with curvature. This can be achieved in more than one way, and below we derive
three possible forms of coupling, all of which confine the systems to the same synchronization manifold, 
but result in different dynamics on this submanifold. 
\begin{figure}[H]
    \centering
    \includegraphics[scale=0.45]{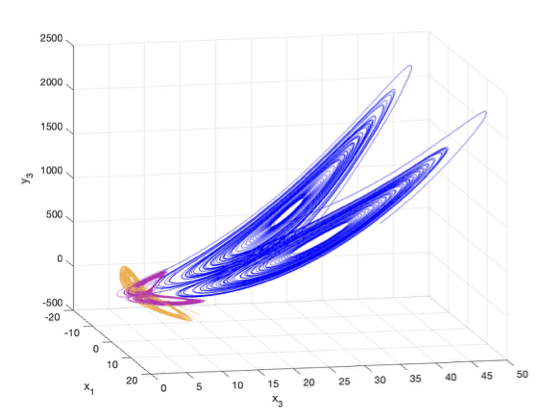}
        \caption{(Colour online) Projection of the dynamics in the coupled system, now confined to the subspace 
        defined by $x_1=y_1, x_2=y_2, x_3=y_3^2$.  The different coupling schemes bring the 
        dynamics to different regions within this submanifold while retaining the characteristics of the two
        oscillators, namely their chaotic nature. The value of $\epsilon$ is 1.
        See text for details.}
    \label{fig:nonlinear}
\end{figure}
\begin{figure}[h]
\centering
    \includegraphics[scale=0.7]{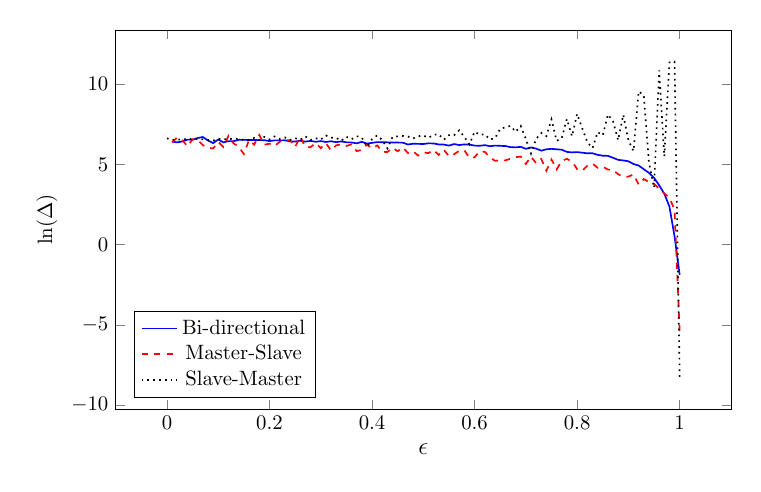}
\caption{(Colour online) The transition to nonlinear projective synchrony using the three different 
coupling forms, as a function of the strength $\epsilon$, as seen in terms of the order parameter 
$\Delta$ defined in Eq.~(\ref{op}). For the 
(i) Master-Slave coupling Eq.~(\ref{nonlinear_ms}) (red dashed line), 
(ii) Slave-Master coupling Eq.~(\ref{nonlinear_sm}) (black dotted line), and 
(iii) bidirectional coupling, Eq.~(\ref{nonlinear_bi}) (solid blue line). Note the 
logarithmic scale on the ordinate. In all three cases the systems show GS 
only for $\epsilon = 1$.}
\label{quaderror}
\end{figure}
The first form of coupling is unidirectional with the master $\x$ subsystem dynamics unaltered, 
forcing the (slave) $\y$ subsystem to modify its behaviour  so as to satisfy the constraints. 
The coupling term $\coupling_1$ is thus a null vector, and the slave coupling function $\coupling_2$ is
\begin{align}
    \coupling_1 = \begin{bmatrix}
0 \\ 0 \\ 0 
\end{bmatrix},
 \quad
\coupling_2 = \epsilon\begin{bmatrix}
0 \\ -x_3 y_1 + x_1 x_3^2 \\ -x_1x_2+2 x_1 x_2 x_3 -\beta x_3^2
\end{bmatrix}, 
\label{nonlinear_ms}
\end{align}
leading to the trajectory coloured blue shown in Fig.~\ref{fig:nonlinear}. 
Alternatively, the $\y$ subsystem could be made the master and $\x$ the slave 
by imposing the condition $x_3= \sqrt{y_3}$. This second form of coupling has 
the advantage of keeping the variables from taking on very large values which may be important 
in a practical  implementation. The coupling (including additional stabilizing terms) for this case is,
\begin{align}
\coupling_1 = \epsilon \begin{bmatrix}
0\\ 
x_1(x_3 - y_3) + (y_2 - x_2) \\ 
-x_1 x_2 + x_1 x_2 x_3/(2 y_3) + \beta x_3/ 2 + (y_3 - x_3^2)
\end{bmatrix}, 
\coupling_2 = \begin{bmatrix}
0 \\ 
0 \\ 
0
\end{bmatrix}.
\label{nonlinear_sm}
\end{align}
This coupling results in the orbit colored magenta in  Fig.~\ref{fig:nonlinear}. 
Finally, we consider bidirectional coupling, in which the systems influence each other; both $x_3$ and $y_3$ adjust 
their values to satisfy the constraints, and one form of such bidirectional coupling that is effective is given by
\begin{align}
    \coupling_1 = \epsilon \begin{bmatrix}
    0 \\ 
    y_1(\rho - y_3) - y_2 + (y_2 - x_2) \\ 
    (y_1 y_2 - \beta y_3) / (2 x_3)
    \end{bmatrix}, 
    \coupling_2 = \epsilon \begin{bmatrix}
    0 \\ 
    x_1(\rho - x_3) - x_2 \\ 
    -2 x_3 (\beta x_3 - x_1 x_2) + (x_3^2 - y_3)
    \end{bmatrix}.
    \label{nonlinear_bi}
\end{align}
This gives the orbit in brown shown in Fig.~\ref{fig:nonlinear}. Note that in the master--slave configuration, 
one of the systems retains the original (or intrinsic) Lorenz dynamics, but with bidirectional coupling, 
the dynamics of both subsystems can be modified while ensuring that the motion occurs on the desired submanifold. 
Since the control objective is algebraic, with other forms of bidirectional coupling the dynamics can be 
drastically altered while keeping the motion on the specified submanifold. Note that
unlike the simple projective synchronization case, here the target submanifold is reached only for $\epsilon$ =1
as shown in Fig.~\ref{quaderror}.\\

\noindent {\bf Translational Constraints:}

The present method becomes particularly simple when the constraint is a translational shift. Consider
two systems in the plane, say,  with coordinates ($x_1, x_2$), and ($y_1, y_2$), governed by the evolution equations 
\begin{eqnarray}
\dot{x}_i &=& v_{x_i}(x_1,x_2)\nonumber\\
\dot{y}_i &=& v_{y_i}(y_1,y_2)~~~~~i=1,2.
\end{eqnarray}
The velocity functions $v$ (with the subscripts indicating the specific variables) determine the motion of the two systems.
If the systems are required to have a fixed separation, given by the conditions $x_i-y_i = a_i$, the 
methodology outlined above -- see Eq.~(\ref{translationalContraintMainEqn}) -- results in master--slave coupling with equations of motion
 \begin{eqnarray}
\dot{x}_i &=& v_{x_i}(x_1,x_2)\nonumber\\
\dot{y}_i &=& v_{x_i}(x_1,x_2) +(x_i-y_i - a_i) ~~~~~i=1,2.
\end{eqnarray}
The slave system has the same dynamics as the master and maintains a specified separation, $(a_1,a_2)$. Applying to the case of two van der Pol oscillators \cite{vdp} with equations of motion
\begin{align}
    \dot{x}_1&= x_2 \nonumber\\
    \dot{x}_2&=\mu x_2(1-x_1^2)-x_1 \nonumber\\
        \dot{y}_1&= y_2  \nonumber\\
    \dot{y}_2&=\mu y_2(1-y_1^2)-y_1
\end{align}
with constraint $x_i-y_i=a_i$,  the equations of motion after coupling, taking the $x$ subsystem as the master 
and the $y$ subsystem the slave, become
 \begin{align}
     \dot{x}_1&= x_2 \nonumber\\
    \dot{x}_2&=\mu x_2(1-x_1^2)-x_1\nonumber\\
     \dot{y}_1&= x_2+x_1-y_1-a_1  &\equiv v_{x_1} +(x_1-y_1-a_1)\nonumber\\
    \dot{y}_2&=\mu x_2(1-x_1^2)-x_1 + (x_2-y_2-a_2)&\equiv v_{x_2} +x_2-y_2-a_2.
    \label{spatial_vdp}
\end{align}

\subsection{Circuit Implementation}

\subsubsection{Projective Synchrony}
Analog realizations of the Lorenz system have been studied in detail for some time now \cite{lorenz_circuit,circuit} 
and there are several ways in which an electronic circuit can be constructed such that the relevant equations are 
identical to Eq.~(\ref{lorenz_eqn}). We utilize $\mu$A741 
operational amplifiers to construct integrator, addition, and multiplication circuits, 
while AD633 is employed for multiplication operations. The resistor values are scaled to 1 megohm, 
and the equations are normalized to 0.1V, resulting in the multiplier 
output being scaled by a factor of 100. The operational amplifiers are biased with $\pm 12V$. \\
\begin{figure}[h]
    \centering
    \includegraphics[scale=0.20]{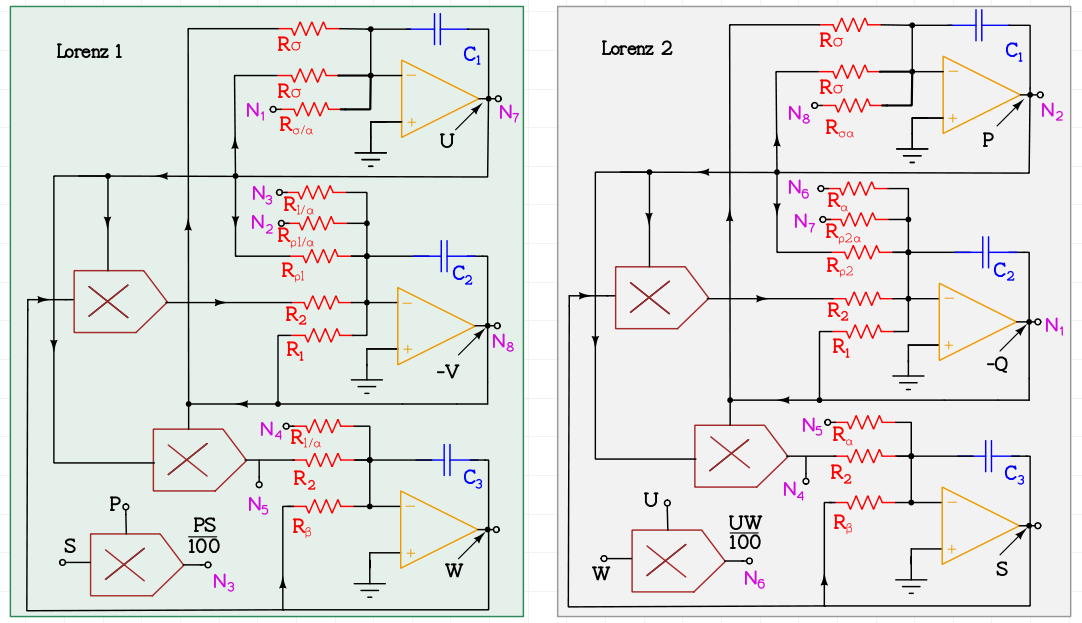}
    \caption{Circuit diagram for the projective synchronization ($x_i-\alpha y_i$ =0), where $\alpha=2.0$. 
    Values of the resistors and capacitors are given in the text, and connections between the two oscillators 
    are shown by the nodes ($N_i$) for simplicity. The respective  paired nodes (say N$_1$-N$_1$) are connected
     during the real-time hardware experiment. }
    \label{fig:case1}
\end{figure}

\begin{figure}[h]
    \centering
    \includegraphics[scale=0.35]{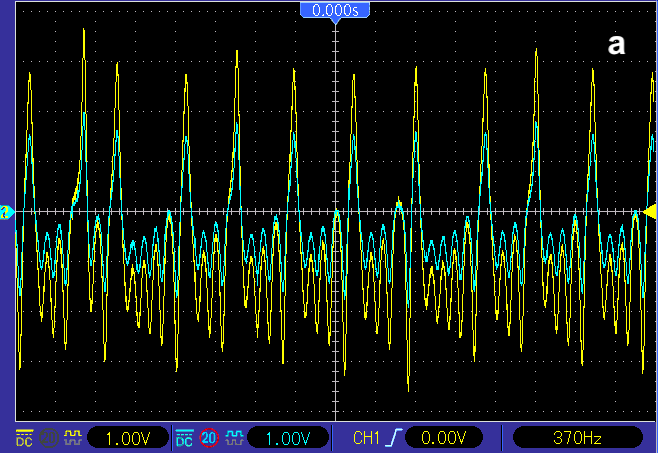}
    \includegraphics[scale=0.35]{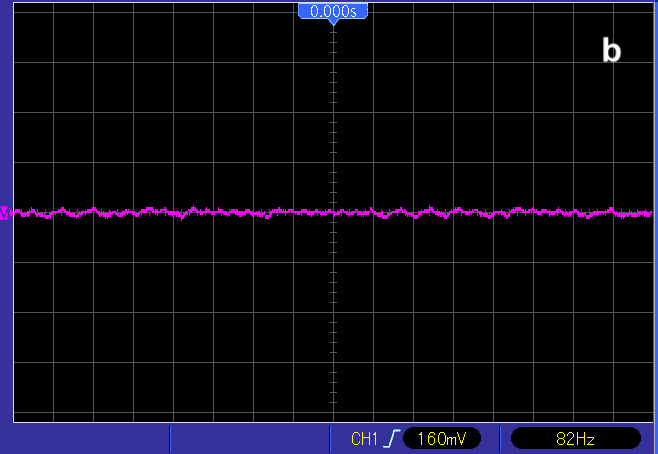}
    \includegraphics[scale=0.35]{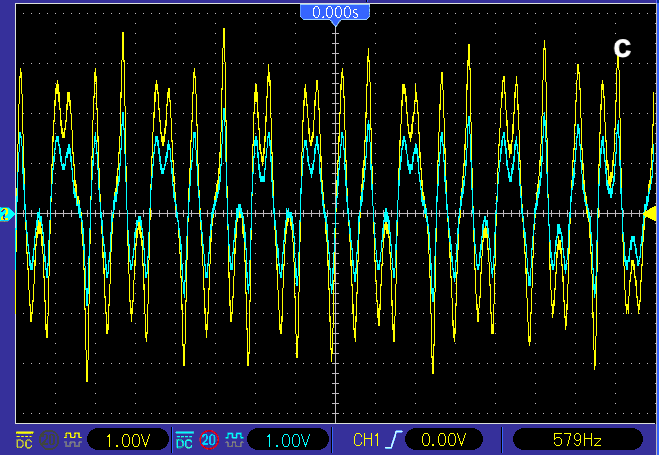}
    \includegraphics[scale=0.35]{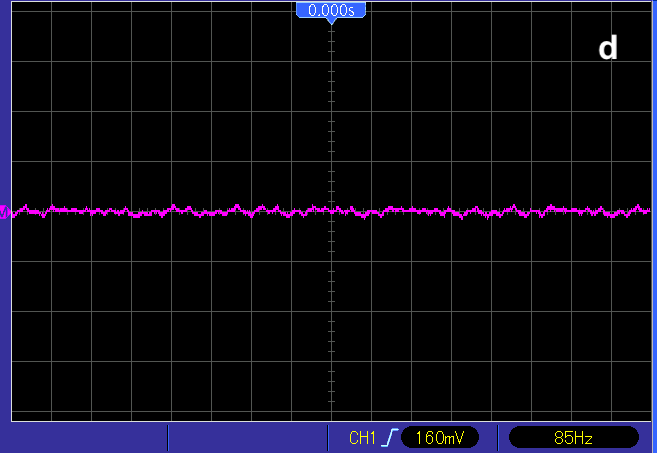}
    
    \caption{Projective synchronization in coupled Lorenz oscillators with system parameters $R_{\rho_y}=35k$, 
    corresponding to a normalized parameter value of $\rho_y=28.57$ in (a). The vertical axis is set 
    at 1V/div for both waveforms. The waveform associated with variable $x_1$ (yellow) exhibits a larger 
    amplitude compared to variable $y_1$. The error ($x_1-\alpha y_1$) is 
    plotted in pane (b). Panels (c) and (d) are for the case $R_{\rho_y}=21k$, namely a normalized $\rho_y$ 
    value of 47.6. The two oscillators maintain the relation $x_1= \alpha y_1$.  The error 
    ($x_1-\alpha y_1$)is plotted in (d)}
    \label{fig:case1_expt1}
\end{figure}

The circuit in Fig.~\ref{fig:case1} corresponds to  Eq. (\ref{bi-directional}), with coupling
strength $\epsilon$ set to 1 and for $\alpha_i =$ 2; details 
can be found in Appendix A. (Results for other choices of $\alpha_i$ are similar). The equations of motion are 
\begin{eqnarray}
\dot{x}_1&=&  \sigma(x_2-x_1)+\sigma y_2/\alpha,\nonumber\\
\dot{x}_2&=&  -x_1x_3+\rho_x x_1-x_2 +\rho_x y_1/\alpha-y_1 y_3/\alpha\nonumber\\
\dot{x}_3&=&  x_1x_2-\beta x_3 +y_1y_2/\alpha, \nonumber\\
\dot{y}_1&=&  \sigma(y_2-y_1) +\sigma \alpha x_2,\nonumber\\
\dot{y}_2&=&  -y_1y_3+\rho_y y_1-y_2+\rho_y \alpha x_1-\alpha x_1 x_3 \nonumber\\
\dot{y}_3&=&  y_1y_2-\beta y_3+\alpha x_1x_2
\label{case1_lorenz_normalised}\end{eqnarray} \noindent

We constructed the circuit (see Eq. (\ref{case1_circuit1}) in Appendix A) on a breadboard with the aforementioned components,
using  AD633JN multiplier ICs,  $\mu$A741 operational amplifiers,  quarter-watt resistors, 
and polyester capacitors with a capacitance of $4.7nF$,
and base resistances chosen to be  $\R=\R_1=1M$, $\R_2={\R}/{100}$, $\R_{1/\alpha}={\R}/{50}=20k$, $\R_\alpha=5k$, 
$\R_{\beta}=347k$, $\R_{\sigma}=100k$, $\R_{\alpha\sigma}=50k$, $\R_{\sigma/\alpha}=200k$, 
$\R_{\rho_x}=35.7k$, $\R_{\rho_y\alpha}=18k$, which corresponds
to parameter values
$\sigma=\R/\R_\sigma=1M/100k=10$, $\rho_x=1M/35.7k=28.0$, $\beta=1M/347k=2.88$, 
$\sigma/\alpha=\R/\R_{\sigma/\alpha}=1M/200K=5.0$,  $\rho_x/\alpha=\R/\R_{\rho/\alpha}
=1M/70.1k=14$,  $\alpha\sigma=\R/\R_{\alpha\sigma}=1M/50k=20$, $\rho_y\alpha=\R/\R_{\rho_y\alpha}=56$.
The output of the circuit was recorded using a 1GSa/s and 100MHz mixed-signal oscilloscope. By considering the specified 
circuit parameters and conditions, we examined the temporal behaviour of the coupled Lorenz circuit. 
The dynamics of the system are depicted in Fig. \ref{fig:case1_expt1}, which shows snapshots of the time series of 
variables $x_1$ and $y_1$. The yellow waveform  corresponds to the circuit variable $\U$, namely $x_1$, while the 
aqua waveform represents the variable $\P$, namely $y_1$.\\

We have studied the synchronized dynamics for several different values of the internal system parameters;
Fig~\ref{fig:case1_expt1}(a)  corresponds to $\R_{\rho_y}=35k$, which translates to a normalized parameter 
value of $\rho_y=28.57$. (Note that the vertical axis is 1V/div for both waveforms.) We have also verified 
that $x_1-\alpha y_1\approx 0$, with small deviations from zero caused by intrinsic circuit noise and 
the inevitable (but small) parameter mismatch.  Fig~\ref{fig:case1_expt1}(b) is the plot of $x_1-\alpha y_1$ 
and in the $y$-axis, the voltage per division is the same as in Fig.~\ref{fig:case1_expt1}(a). 
Each resistor or capacitor has an inherent tolerance that affects its actual value.\\ 

In the present circuits we use several multipliers (AD633JN) and op-amps ($\mu A741$). 
Each multiplier is responsible for performing operations such as ${x \cdot y}/{10}$ or ${x \cdot y}/{100}$, 
each of which may introduce up to $2\%$ error (as specified for the performance at 25°C with a 2 k$\Omega$ 
output load). Further, the error accumulates with sequential multiplications. Similar considerations 
apply to the op-amps used in the circuit, and these introduce other tolerance-related errors. In the 
projective synchronization wherein we had set $\alpha = 2$, the experimental data gives, on average,  about 
$8\%$ deviation from the ideal values.  Circuit components such as resistors, capacitors, multipliers, 
and op-amps are the primary causes of this deviation.\\

We have studied projective synchronization for another value of the system parameter,
$\R_{\rho_y}=21k$ corresponding to a normalized $\rho_y$ value of 47.6. As discussed earlier, the two oscillators 
maintain the relation $x_1=\alpha y_1$ (Fig.~\ref{fig:case1_expt1}(c)); the error can be seen in 
Fig.~\ref{fig:case1_expt1}(d). When the parameters of the master Lorenz system is dynamically
modified, the dynamics of the other system adjusts accordingly to maintain synchronization on the 
manifold $x_i-\alpha y_i=0$. 

\subsubsection{Nonlinear Scaling}

\begin{figure}[h]
    \centering
    \includegraphics[scale=0.3]{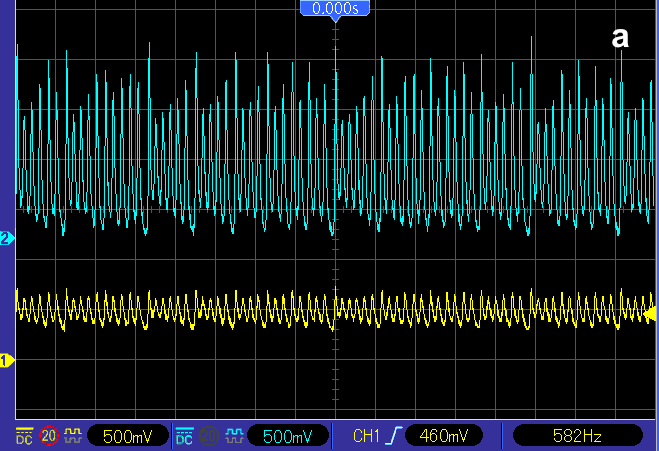}
    \hspace{1cm} 
    \includegraphics[scale=0.3]{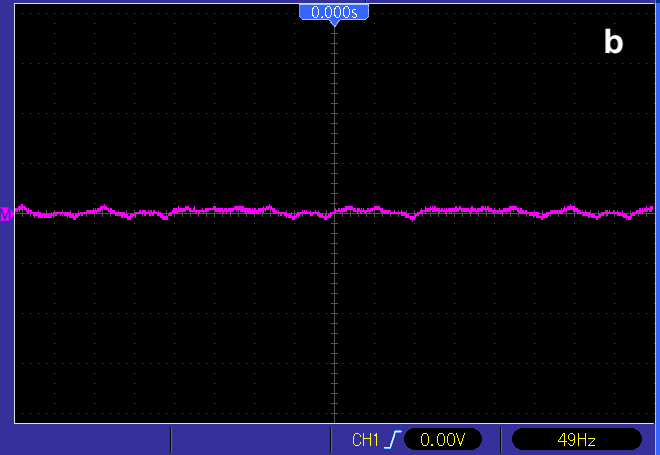}\\
    \includegraphics[scale=0.8]{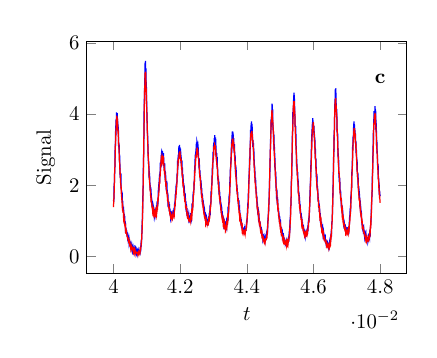}
    \includegraphics[scale=0.8]{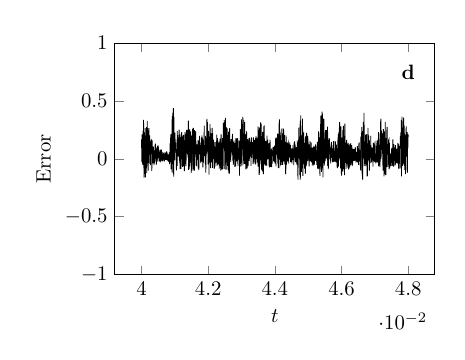}
    \caption{Generalized synchronization of coupled Lorenz oscillators with 
    nonlinear scaling for (a) parameter $R_{\rho_y}=34.48k$ ($\rho_y=29$). 
    The yellow waveform corresponds to variable $x_3$ and the aqua waveform represents variable $y_3$ and panel
    (b) shows the error ($y_3- x_3^2\approx 0$) in the coupled systems. 
    In (c) the parameter is $R_{\rho_y}=21k$ ($\rho_y=47.6$) and the variables depicted 
    are $x_3^2$ (blue) and $y_3$ (red). The relative error is shown in (d).}
    \label{fig:case2_expt}
\end{figure}

The second example we consider is the case $y_1 = x_1$, $y_2 = x_2$, $y_3 = x_3^2$ for two coupled Lorenz systems, 
using the coupling function described in Eq. \eqref{nonlinear_ms}, also with $\epsilon = 1$. We construct the circuit 
shown in Fig.~\ref{fig:case2}, and following the procedure described in Appendix A for the  
variables  U, V, W, P, Q, and S, one obtains, in a straightforward manner, the dynamical equations
\begin{eqnarray}
\dot{x}_1&=&  \sigma(x_2-x_1),\nonumber\\
\dot{x}_2&=&  -x_1x_3+\rho_x x_1-x_2, \nonumber\\
\dot{x}_3&=&  x_1x_2-\beta x_3 ,\nonumber\\
\dot{y}_1&=&  \sigma(y_2-y_1) \nonumber\\
\dot{y}_2&=&  -y_1y_3+\rho_y y_1-y_2 -x_3 y_1 + x_1 x_3^2\nonumber\\
\dot{y}_3&=&  y_1y_2-\beta y_3 -x_1x_2+2 x_1 x_2 x_3 -\beta x_3^2
 \label{case2_lorenz_normalised}
\end{eqnarray}
where resistances were chosen as $\R=2M\Omega$, $\R_3=\R_{\beta}/{100}$, and $\R_4=\R/200$ so that the parameters
become $\sigma=\R/\R_{\sigma}=2M/200k=10$, $\rho_x=\R/\R_{\rho 1}=2M/70.7k=28.36$ and 
$\rho_y=\R/\R_{\rho 2}=2M/70.3k=28.44$, $\beta=\R/\R_{\beta}=2M/650k=3.07$.  Note that we use
 a different combination of parameters in this case since our target objective is that $y_3 = x_3^2$. To ensure that the circuit 
 oscillation remains well below the saturation/operating voltage of the op-amp, we scaled the circuit accordingly.  The parameters for the two uncoupled Lorenz oscillators are carefully chosen to exhibit chaotic dynamics. With coupling, the system exhibits GS  with the specific relation $y_3=x_3^2$; see Fig.~\ref{fig:case2_expt}. \\
\begin{figure}[h]
    \centering
    \includegraphics[scale=0.30]{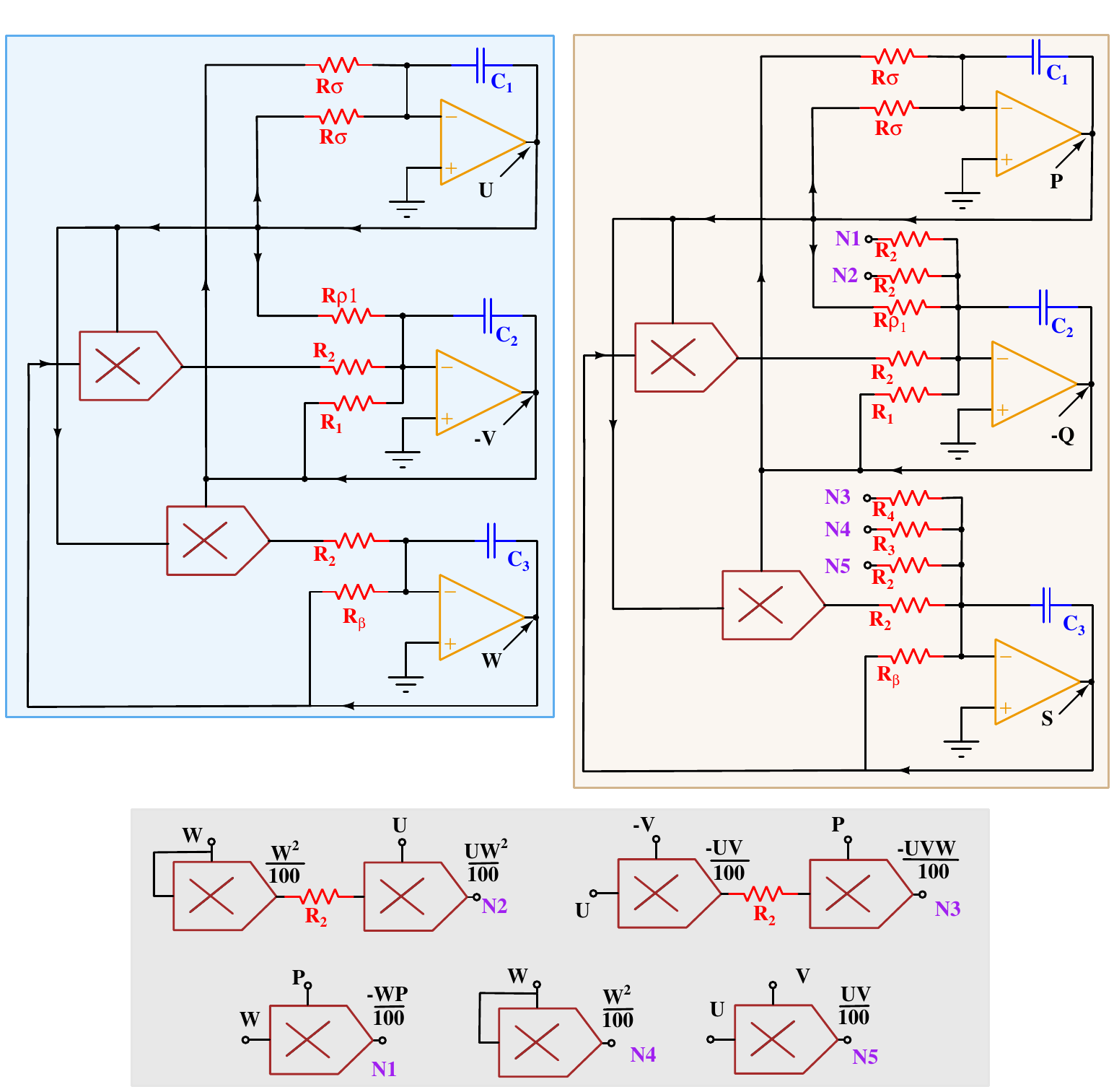}
    \caption{Circuit diagram for the nonlinear scaling ($y_3 = x_3^2$). Two Lorenz oscillators are shown in separate boxes.  
    Values of the resisters and capacitors are given in the text. Connection between the two oscillators are shown by the 
    nodes (Ni) for simplicity. The respective  paired nodes (say N$_1$-N$_1$) are connected during the real-time hardware experiment. }
    \label{fig:case2}
\end{figure}

The relationship between the two signals, $x_3$ and $y_3$ is shown in Fig.~\ref{fig:case2_expt}.
Fig.~\ref{fig:case2_expt}(a) is for $\R_{\rho_y}=34.48k$, which corresponds to a 
normalized parameter value of $\rho_y=29$. In the snapshot, the vertical axis ($y$-axis) 
is set at 500mV/div for both waveforms. The waveform in 
yellow represents the variable $x_3$, while the aqua waveform corresponds to the variable $y_3$. 
Fig.~\ref{fig:case2_expt}(b) shows the error ($y_3- x_3^2\approx 0$)and this can  also be 
verified using the recorded data.
The time-series shown in Fig.~\ref{fig:case2_expt}(c) is for $R_{\rho_y}=21k$, the normalized 
value of $\rho_y$ being 47.6. The variable $x_3 ^2$ is shown overlaid on the $y_3$ time series,
and as can be see, the relative error is quite low. Given the 
tolerances of the off-the-shelf components, the maximum error remains below $\approx
5\%$ as shown in Fig.~\ref{fig:case2_expt}(d).

\subsubsection{Translational Constraints} \label{circuitTranslationalConstrain}

\begin{figure}[h]
    \centering
    \includegraphics[scale=0.45]{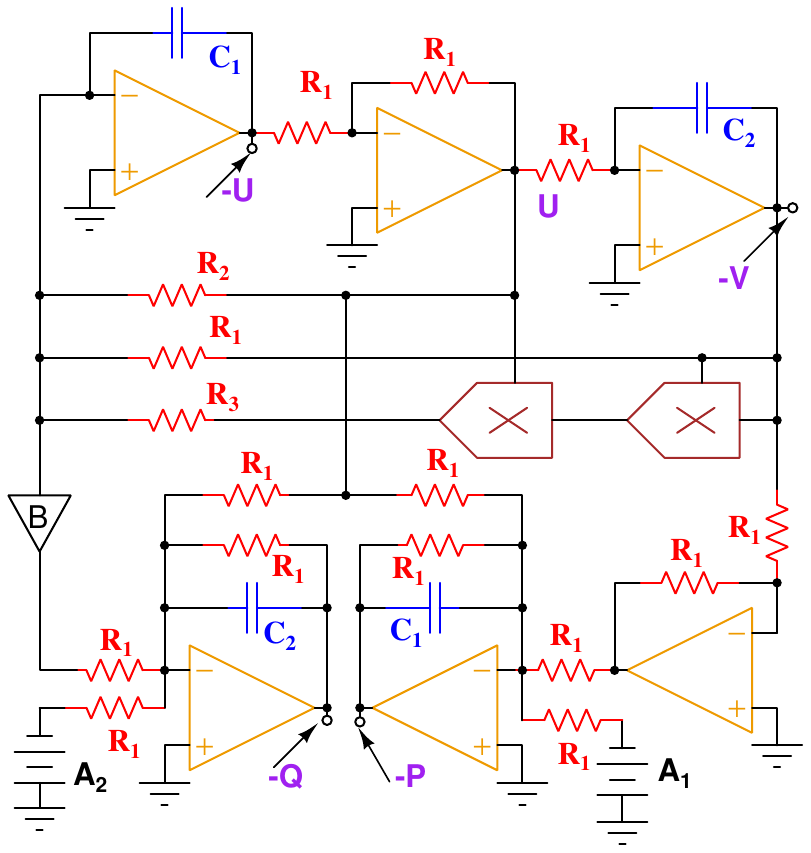}
    \caption{Circuit diagram for two coupled van der Pol oscillators designed to maintain a 
    translational separation as discussed.  The DC voltage sources corresponding to the 
    constants $a_1$ and $a_2$ are given by $\A_1$ and $\A_2$ and the buffer circuit is marked as B.}
    \label{fig:case3}
\end{figure}

 \begin{figure}[h]
    \centering
    \includegraphics[scale=0.3]{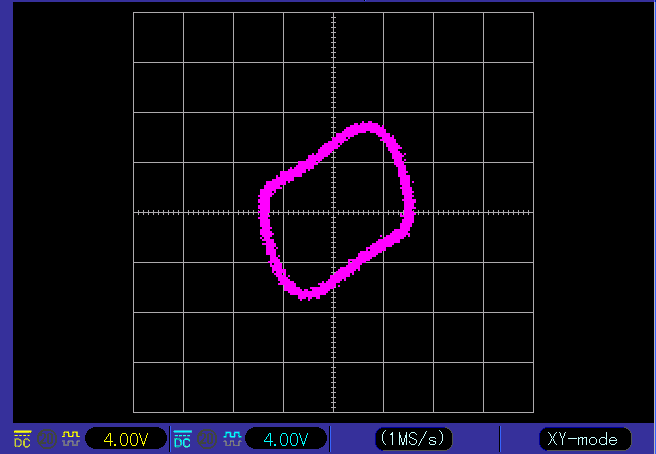}
    \includegraphics[scale=0.3]{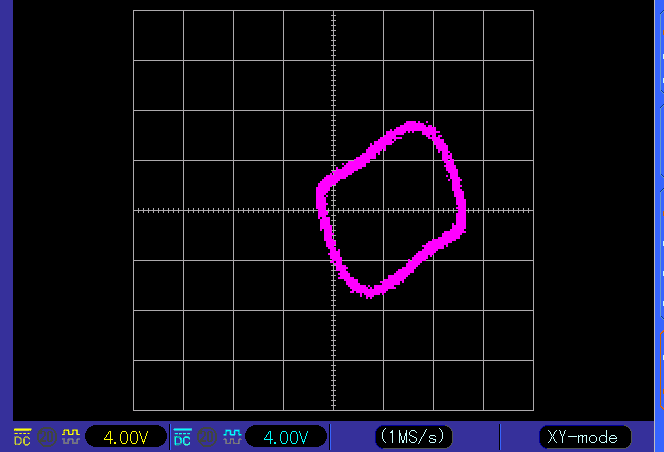}\\
    \includegraphics[scale=0.3]{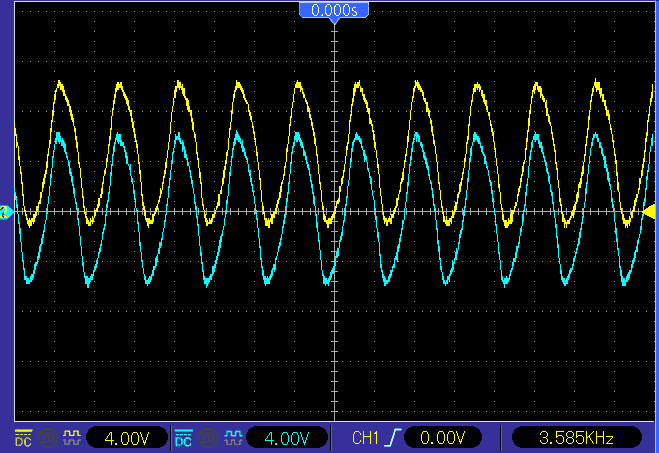}
    \includegraphics[scale=0.3]{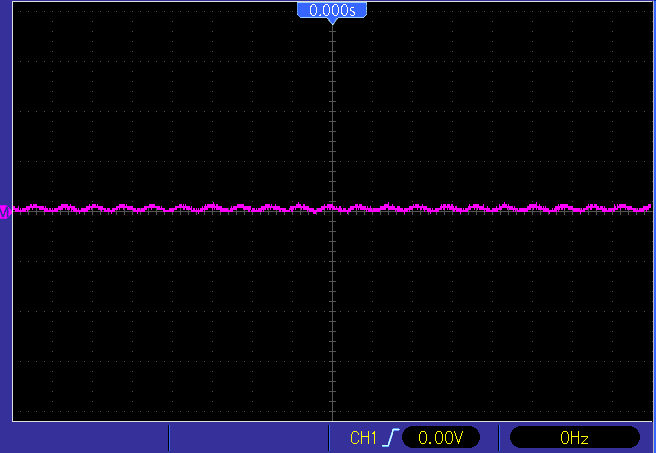}\\
    \caption{Parameters were adjusted so that $a_1$=5, and $a_2$=0. 
 The dynamics of the master system are shown in (a) which plots the variables ($\V, \U$), with the  origin at (0V, 0V). 
 The dynamics of the slave system is shown in (b) which plots the variables ($\Q, \P$ which are offset by (5V, 0V). 
 There is some saturation in the $x$-axis of the slave system when the circuit voltage reaches the saturation 
 voltage of the op-amp. 
The synchronized behaviour between the master and slave systems with the translation shift can be seen in (c) which plots
 $\V$ (aqua) and $\Q$ (yellow) as a function of time.  (d) shows the error between of the two systems ($x_i-y_i - a_i\approx 0$), we also used recorded data to verify the relationship.}
    \label{fig:case3_expt}
\end{figure}

The corresponding circuit  is shown in Fig.~\ref{fig:case3}, 
where the circuit parameters are $\R=\R_1=10k\Omega, \R_2=20k\Omega, \R_3={\R_2}/{100}$, and $\mu={\R}/{\R_2}=0.5$. 
The buffer circuit is marked B in  Fig.~\ref{fig:case3}. The DC voltage  sources corresponding to the constants 
$a_1$ and $a_2$ are given by $\A_1$ and $\A_2$. \\

Experimental snapshots of the evolution of the coupled van der Pol oscillator system are captured on the oscilloscope and are 
shown in Fig. \ref{fig:case3_expt}. In Fig. \ref{fig:case3_expt}(a), the snapshot displays the dynamics of the master system, 
specifically the variables ($\V,\U$). The master system has its origin at (0V,0V). On the other hand, 
Fig. \ref{fig:case3_expt}(b) illustrates the dynamics of the slave system, observed through the variables ($\Q,\P$). 
The slave system is spatially shifted to (5V,0V), as can be clearly seen. A slight saturation in the $x$-axis occurs 
when the circuit voltage reaches the saturation voltage of the op-amp. Interestingly, despite the translational separation 
between the master and slave systems, they exhibit synchronization, most clearly seen in Fig. \ref{fig:case3_expt}(c) 
that showcases the system variables $\V$ (aqua) and $\Q$ (yellow), representing the master and slave systems, respectively. Fig. \ref{fig:case3_expt}(d) is the snapshot of error between of the two systems ($x_i-y_i - a_i\approx 0$).\\

This can be extended to more dimensions and for more than two systems, and forms the basis of an effective 
swarming algorithm (see Section~\ref{swarmDronesSectionTitle}) that has similarities to methods that 
have been devised for collision avoidance of aircraft in a convoy \cite{patent}.
\section{Swarming} \label{swarmDronesSectionTitle}

Translational coupling between the systems as discussed above provides a basis for swarming.
A centralised leader-follower based swarming algorithm, where one system serves as the leader 
(or master) and the remaining systems act as followers (or slaves) is achieved by imposing 
translational constraints on the followers relative to the leader and maintaining this 
constraint at all times. 

To illustrate this, we consider coupling the dynamics in $x$- and $y$- directions, 
with no coupling in the $z$-direction.
If $\dot{x}_m$ and $\dot{y}_m$ are the velocities of the leader 
\begin{align}
\left.
\begin{array}{rl}
\dot{x}_m &= v_{x_m}(x_m)\nonumber\\
\dot{y}_m &= v_{y_m}(y_m)\nonumber
\end{array}
\right\} &\text{ Leader system}
\end{align}

and $\dot{x}_i$ and $\dot{y}_i$ represent the velocities of the $i^\text{th}$ follower, 
with $a_i$ and $b_i$ the defined separations (or constraints), then Eq.~(\ref{swarming_eqn}) 
will ensure that the follower maintain a defined separation concurrently following the motion of the leader. 
\begin{align}
\left.
\begin{array}{rl}
\dot{x}_i &= v_{x_m}(x_m) + (x_m - x_i - a_i)\\
\dot{y}_i &= v_{y_m}(y_m) + (y_m - y_i - b_i)
\end{array}
\right\} &\text{ Follower system} \quad i = 1, 2, \ldots, n
\label{swarming_eqn}
\end{align}

In the supplementary material, a movie simulation of one leader and 24 followers following an arbitrary
trajectory in three dimensions is given as SW.mp4.\\

We implement this method for a set of drones to test its robustness on real-world systems. 
A set of five drones were used in the experiment, namely one leader and four follower drones. 
The drones operates on the ArduPilot software which incorporates the Python package {\tt Pymavlink}
for control and communication. Each drone is equipped with a built-in GPS and connected to a computer 
via a WiFi router. These drone uses NED (or North East Down) frame for the position and we therefore
converted this to a common local frame in which the origin of the frame was at the starting position of the 
leader drone. A swarming script running on the computer captured the GPS coordinates from the drones, 
computed their subsequent positions, which were transmited back to the drones. This script functioned
continuously and iteratively, ensuring communication between the drones at a frequency of 20 Hz. 
The path of the leader has been decided in real-time by providing the GPS coordinates.

\begin{algorithm}
\caption{Swarming Algorithm Script Flow}
\begin{algorithmic}[1]
\State \textbf{Initialize} UDP input ports for each drone and set desired separations $(a_i, b_i)$ for all follower drones
\For{each drone $i$}
    \If{heartbeat signal is \texttt{True}}
        \State Set flight mode to \texttt{GUIDED}
        \State Arm the drone and initiate takeoff to a target altitude of 10 meters
    \EndIf
\EndFor
\State Receive initial positions $(x_i, y_i)$ of all drones at the Global Control Station (GCS)
\While{heartbeat signals from all drones are \texttt{True}}
    \State Move the leader drone as desired, either manually via radio control or programmatically through the GCS
    \State Convert leader's global coordinates to the local frame
    \State Compute the desired positions of all follower drones using Eq.~\eqref{swarming_eqn}
    \State Transform these local-frame positions back to NED frame
    \State Transmit the updated GPS positions to each follower drone for navigation
\EndWhile
\State Store operation data and flight logs for post-analysis
\end{algorithmic}
\label{swarmingAlgorithm}
\end{algorithm}

The experiment was conducted in an open field at 28.54424 N, 77.18847 E under ambient conditions, at an
altitude of 10 m. The spatial separations between each of the followers and the master drone are constant
and the values used were (in meters):
\begin{equation*}
\begin{aligned}
(a_1, b_1) &=\; ( \phantom{-}3, -3) \\
(a_2, b_2) &=\; ( \phantom{-}3,  \phantom{-}3) \\
(a_3, b_3) &=\; (-3, -3) \\
(a_4, b_4) &=\; (-3,  \phantom{-}3)
\end{aligned}
\end{equation*}

\begin{figure}[h]
    \centering
    \includegraphics[scale=0.55]{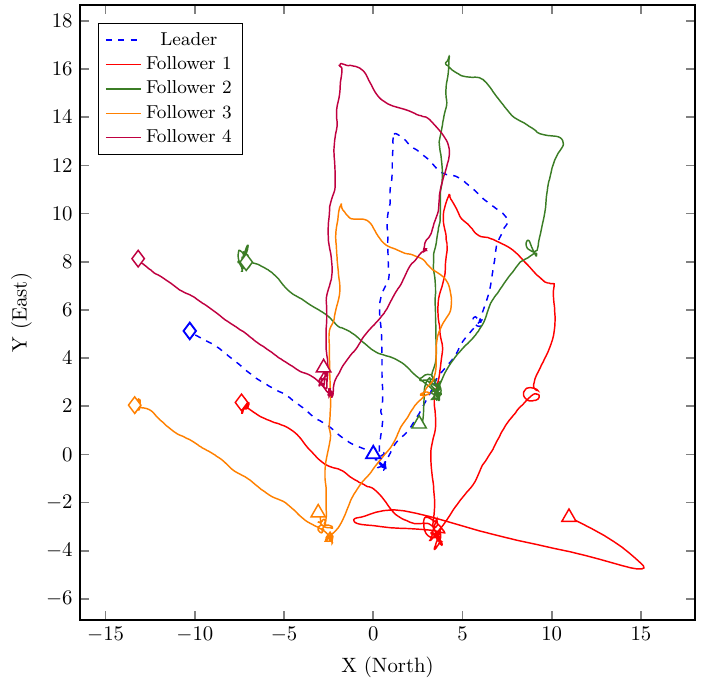}
\caption{Dynamics of the master drone (dashed blue line) and other four slave drones. The plot 
shows the path covered by each of the drones in the field. The $x$-axis points north and the $y$-axis 
points east, and the starting point of the leader drone is taken as the origin. The triangle 
marker represents the starting point of the drones and the diamond markers are the ending positions. 
A movie of the drones in action is given in the Supplementary Material, Fig.11.mp4}
    \label{fig:timeIndPath}
\end{figure}

\begin{figure}[h]
    \centering
    \includegraphics[scale=0.5]{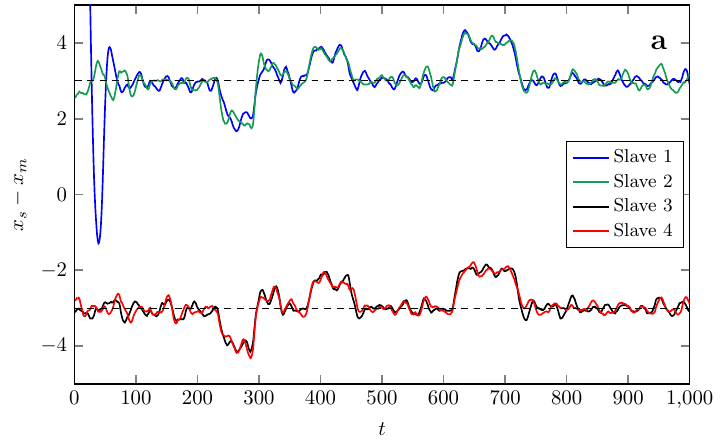}
    \includegraphics[scale=0.5]{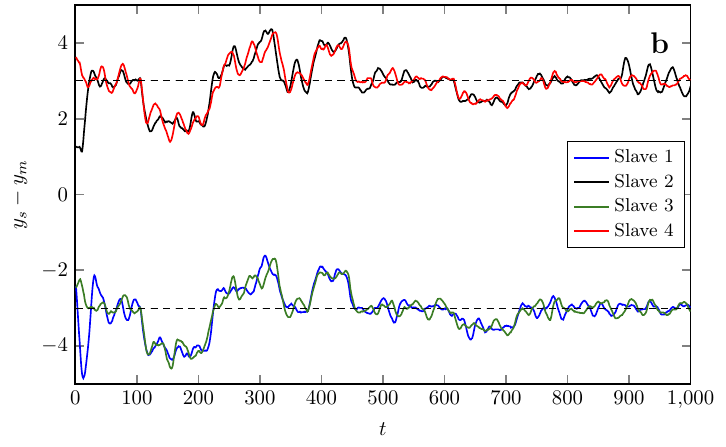}
    \includegraphics[scale=0.5]{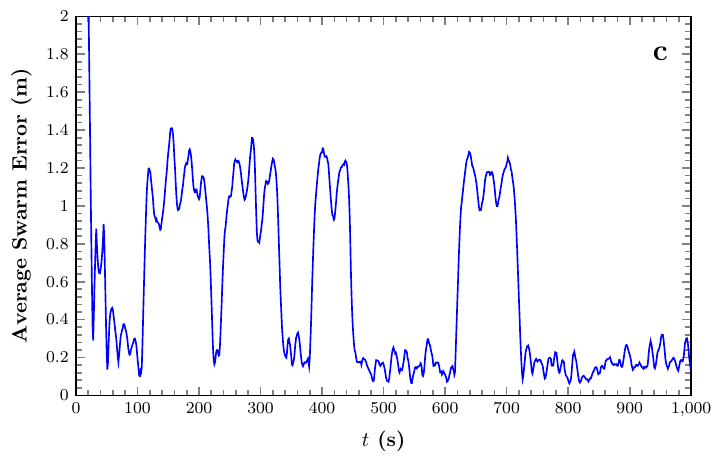}
\caption{real-time data from the drones for the case of fixed (time-independent) separations. 
(a) Displacement in the $x$ direction between master and slave drones vs time. (b) Displacement in the $y$ direction 
between master and slave drones vs time. \textcolor{red}{(c) The average error of the swarm (defined in the text) 
as a function of time. Note the difference in scale.}}
\label{fig:timeIndTimeSeries}
\end{figure}

Drone trajectories are shown in Fig.~\ref{fig:timeIndPath} and it can be seen that the follower drones 
quickly align with the master, while simultaneously maintaining the defined separation. Details of the time-series 
are given in Fig.~\ref{fig:timeIndTimeSeries}(a) and Fig.~\ref{fig:timeIndTimeSeries}(b). 
\textcolor{red}{The average error is equivalent to the average synchronization error in this case,
namely $E_{\mathrm{avg}}(t) = \frac{1}{N} \sum_{i=1}^{N} \left\lVert \; | \mathbf{x}_i(t) - \mathbf{x}_m(t)| -\mathbf{a}_i\right\rVert$, 
where $\mathbf{x}$ \& $\mathbf{a}$ are vectors representing drone positions and required 
separation respectively. When there are sudden changes in the direction of the leader the 
followers have a lag of a few seconds: this leads to a spike in the error, but the system 
stabilizse rapidly. This can be seen in Fig.~\ref{fig:timeIndTimeSeries}(c) where the change 
in the direction of the master is correlated to the surges in the average error.} 
The minor deviations are primarily caused by environmental noise 
such as wind which are managed by the onboard PID controller for each of the drone. The resistance 
to perturbation and maintenance of a fixed position are strongly influenced by the PID controller 
and the design of the drone. 

\section{Discussion and Summary}

In this paper we have demonstrated the practical implementation of a control method to achieve 
specific forms of generalized synchronization in coupled nonlinear systems \cite{CR}. The control
is achieved through imposition of constraints: when dynamical systems are in synchrony, the combined
dynamics occurs on a submanifold in the phase space of the coupled system. Our method designs the
coupling through which these constraints are maintained; this is a method of `synchronization 
engineering' \cite{KISS}. While the primary focus of our work is generalised synchronization, the 
methods we use also offer us some insight into coupling mechanisms, an area of considerable interest 
\cite{SPMS, SPMS2019}. {\color{red} Further, the stability of the control method can be ensured by the addition of 
suitably designed stabilizers that encode appropriate Lyapunov functions, as discussed in Section 2.}\\

Pairs of electronic circuits corresponding to the chaotic Lorenz and the van der Pol oscillators 
were constructed and coupled appropriately so that the variables of one system have a specified relationship with
those of the other. Examples studied here included cases of linear and nonlinear scaling. Since our methods has considerable
flexibility, a variety of couplings can be designed in order to target a given generalized synchronization objective; this allows 
us to use couplings that minimally alter the dynamics of the interacting systems. An additional advantage is that one can design
coupling terms that can be physically realized in a given situation (for instance, not every algebraic form of interaction can be
translated into off-the-shelf circuit components).\\

One of the simplest controls that can be implemented is a translational one, when the variables of the two systems 
are related by a constant shift. This makes it possible to achieve spatiotemporal synchrony, which is particularly 
useful in understanding swarming and flocking behaviour. The application here consisted of the concerted movement 
of a set of drones, one of which was treated as the master from which the remaining drones were constrained to 
move at a specified distance. The resulting swarm-like behaviour was satisfactory. 
Our procedure has some similarity to the dynamics of the Vicsek model \cite{vicsek}: the velocities are  
effectively identical, and since one of the systems is assigned the role of a leader, the translational
constraints serve as a means of aligning the orientations, so that the ensemble moves as a swarm. \\

Extensions of this technique to incorporate time-dependent separations between the leader and followers 
are currently being investigated. \textcolor{red}{A straightforward implementation of 
sinusoidally varying distance constraint was tested experimentally in a drone swarm 
and we found that to a good approximation, the slave systems obey the constraint. In the general 
time-dependent case though, the coupling functions will also depend on the rate of 
variation of the constraint and the control becomes correspondingly more complicated. 
This requires further experimentation. }

\section*{Acknowledgements} We thank Prof. Soumitro Banerjee of IISER Kolkata for a conversation  
many years ago on implementation of these ideas, and him and Tania Ghosh, then also at IISER Kolkata, 
for doing some initial circuit simulations 
of the basic projective synchronization.  During the period when this research was initiated, hospitality of the
Department of Chemistry, IIT Delhi and the support of the SERB, India in the form of a J C Bose Fellowship to 
RR is gratefully acknowledged. 
SK acknowledges financial support from the Center for Computational Modelling, Chennai Institute of 
Technology, India, under the grant CIT/CCM/2024/RP/005.

\appendix
\section{Circuit equations}
It is straightforward to see that the circuit in  Fig.~\ref{fig:case1} corresponds to the bidirectionally coupled Lorenz oscillator system 
in Section \ref{PS}, namely Eqs.~(\ref{linear_coupling}), with $\alpha_i =$ 2.  (Results for other choices of $\alpha_i$ are similar.) 
The output transfer function of the circuit at $U, V, W, P, Q$ and $S$ is represented as
\begin{eqnarray}
\U&=& -\frac{1}{\C} \int \left(\frac{\U}{\R_{\sigma}}-\frac{\V}{\R_{\sigma}}-\frac{\Q}{\R_{\sigma/\alpha}}\right) dt, \nonumber\\
\V&=& -\frac{1}{\C} \int \left(\frac{\W\U}{100\R_{2}}-\frac{\U}{\R_{\rho_x}}+\frac{\V}{\R_1}-\frac{\P}{\R_{\rho_x/\alpha}} +\frac{\P\S}{100\R_{1/\alpha}}\right) dt, \nonumber\\ 
\W&=& -\frac{1}{\C} \int \left(-\frac{\U\V}{100\R_2}+\frac{\W}{\R_{\beta}} -\frac{\P\Q}{100\R_{1/\alpha}} \right) dt, \nonumber\\\nonumber\\
\P&=& -\frac{1}{\C} \int \left(\frac{\P}{\R_{\sigma}}-\frac{\Q}{\R_{\sigma}}-\frac{\V}{\R_{\sigma\alpha}}\right) dt, \nonumber\\
\Q&=& -\frac{1}{\C} \int \left(\frac{\S\P}{100\R_{2}}-\frac{\P}{\R_{\rho_y}}+\frac{\Q}{\R_1}  -\frac{\U}{\R_{\rho_y \alpha}}   +\frac{\U\W}{100\R_{\alpha}} \right) dt, \nonumber\\
\S&=& -\frac{1}{\C} \int \left(-\frac{\P\Q }{100\R_2}+\frac{\S}{\R_{\beta}} -\frac{\U\W}{100\R_{\alpha}}\right) dt.
\label{case1_circuit1}
\end{eqnarray} 
Differentiating Eqs. (\ref{case1_circuit1}) with respect to time followed by rescaling each equation by the resistance R
and rearranging, we obtain 
\begin{eqnarray}
\R\C\frac{d\U}{dt}&=&  -\frac{\R}{\R_{\sigma}}\left(\U-\V\right)-\frac{\R}\R_{\sigma/\alpha}\Q, \nonumber\\
\R\C\frac{d\V}{dt}&=&   -\left(\frac{\R}{100\R_{2}}\W\U-\frac{\R}{\R_{\rho_x}}\U+\frac{\R}{\R_1}\V  -\frac{\R}{\R_{\rho_x/\alpha}}\P +\frac{\R}{100\R_{1/\alpha}}\P\S  \right), \nonumber\\
\R\C\frac{d\W}{dt}&=&   -\left(-\frac{\R}{100\R_{2}}\U\V+\frac{\R}{\R_{\beta}}\W -\frac{\R}{100\R_{1/\alpha}} \P\Q\right),\nonumber\\
\R\C\frac{d\P}{dt}&=&  -\frac{\R}{\R_{\sigma}}\left(\P-\Q\right)-\frac{\R}\R_{\sigma\alpha}\V, \nonumber\\
\R\C\frac{d\Q}{dt}&=&   -\left(\frac{\R}{100\R_{2}}\P\S-\frac{\R}{\R_{\rho_y}}\P+\frac{\R}{\R_1}\Q     -\frac{\R}{\R_{\rho_y\alpha}}\U +\frac{\R}{100\R_{\alpha}}\U\W  \right), \nonumber\\
\R\C\frac{d\S}{dt}&=&   -\left(-\frac{\R}{100\R_2}\P\Q+\frac{\R}{\R_{\beta}}\S    -\frac{\R}{100\R_{\alpha}} \U\V \right).
\end{eqnarray}
Rescaling time $t\to t/\R\C$ and making the identification $\U = x_1$, $\V = x_2$, $\W = x_3$, 
$\P = y_1$, $\Q = y_2$, and $\S = y_3$, we obtain the normalized equations corresponding to the 
bidirectionally coupled Lorenz oscillators with the required
constraint $x_i=\alpha y_i$, $i$=1, 2, 3,
\begin{eqnarray}
\dot{x}_1&=&  \sigma(x_2-x_1)+\sigma y_2/\alpha,\nonumber\\
\dot{x}_2&=&  -x_1x_3+\rho_x x_1-x_2 +\rho_x y_1/\alpha-y_1 y_3/\alpha\nonumber\\
\dot{x}_3&=&  x_1x_2-\beta x_3 +y_1y_2/\alpha, \nonumber\\
\dot{y}_1&=&  \sigma(y_2-y_1) +\sigma \alpha x_2,\nonumber\\
\dot{y}_2&=&  -y_1y_3+\rho_y y_1-y_2+\rho_y\alpha x_1-\alpha x_1 x_3 \nonumber\\
\dot{y}_3&=&  y_1y_2-\beta y_3+\alpha x_1x_2
\end{eqnarray} 
\noindent
where the base resistances are  $\R=\R_1=1M$, $\R_2=\R/100$, $\R_{1/\alpha}=\R/50=20k$, $\R_\alpha=5k$, 
$\R_{\beta}=347k$,$\R_{\alpha\sigma}=50k,$ $\R_{\sigma/\alpha}=200k$, $\R_{\rho_y\alpha}=18k$, 
$\sigma=\R/\R_{\sigma}=1M/100k=10$, $\rho_x=\R/\R_{\rho_x}=1M/35.7k=28.0$, $\beta=1M/347k=2.88$, 
$\sigma/\alpha=\R/\R_{\sigma/\alpha}=1M/200K=5.0$,  $\rho_x/\alpha=\R/\R_{\rho/\alpha}
=1M/70.1k=14$,  $\alpha\sigma=\R/\R_{\alpha\sigma}=1M/50k=20$, $\rho_y\alpha=\R/\R_{\rho_y\alpha}=56$.\\

Analysis of the other coupled circuits considered in this paper is similar and straightforward.


\end{document}